\documentclass[aps,prl,twocolumn,showpacs,preprintnumbers,superscriptaddress,nofootinbib]{revtex4-1}
\usepackage{amsmath}
\usepackage{amssymb}
\usepackage{graphicx}
\usepackage{dcolumn}
\usepackage{bm}
\usepackage[usenames,dvipsnames]{color}
\newcommand{\CMS}{\rm CaMn$_{2}$Sb$_{2}$}

\begin{document}

\title{CaMn$_{2}$Sb$_{2}$: Spin waves on a frustrated antiferromagnetic honeycomb lattice}%

\author{D.E. McNally}
\email{daniel.mcnally@stonybrook.edu}
\affiliation{Department of Physics and Astronomy, Stony Brook University, Stony Brook, New York 11794-3800, USA}
\author{J.W. Simonson}
\affiliation{Department of Physics and Astronomy, Stony Brook University, Stony Brook, New York 11794-3800, USA}
\author{J.J. Kistner-Morris}
\affiliation{Department of Physics and Astronomy, Stony Brook University, Stony Brook, New York 11794-3800, USA}
\author{G.J. Smith}
\affiliation{Department of Physics and Astronomy, Stony Brook University, Stony Brook, New York 11794-3800, USA}
\author{J.E. Hassinger}
\affiliation{Department of Physics and Astronomy, Stony Brook University, Stony Brook, New York 11794-3800, USA}
\author{L. DeBeer-Schmitt}
\affiliation{Spallation Neutron Source, Oak Ridge National Laboratory, Oak Ridge, TN 37831-6473, USA}
\author{A.I. Kolesnikov}
\affiliation{Spallation Neutron Source, Oak Ridge National Laboratory, Oak Ridge, TN 37831-6473, USA}
\author{I.A. Zaliznyak}
\affiliation{Condensed Matter Physics and Materials Science Department, Brookhaven National Laboratory, Upton, New York, 11973-5000, USA}
\author{M.C. Aronson}
\affiliation{Department of Physics and Astronomy, Stony Brook University, Stony Brook, New York 11794-3800, USA}
\affiliation{Condensed Matter Physics and Materials Science Department, Brookhaven National Laboratory, Upton, New York, 11973-5000, USA}

\date{\today}%

\begin{abstract}
We present inelastic neutron scattering measurements of the antiferromagnetic insulator CaMn$_{2}$Sb$_{2}$, which consists of corrugated honeycomb layers of Mn. The dispersion of magnetic excitations has been measured along the \textbf{H} and \textbf{L} directions in reciprocal space, with a maximum excitation energy of $\approx$ 24 meV. These excitations are well described by spin waves in a Heisenberg model, including first and second neighbor exchange interactions, J$_{1}$ and J$_{2}$, in the Mn plane and also an exchange interaction between planes. The determined ratio J$_{2}$/J$_{1}$ $\approx$ 1/6 suggests that {\CMS} is the first example of a compound that lies very close to the mean field tricritical point, known for the classical Heisenberg model on the honeycomb lattice, where the N\'eel phase and two different spiral phases coexist. The magnitude of the determined exchange interactions reveal a mean field ordering temperature $\approx$ 4 times larger than the reported N\'eel temperature T$_{N}$ = 85 K, suggesting significant frustration arising from proximity to the tricritical point. 
\end{abstract}


\maketitle

Frustration occurs in spin systems when constraints prevent the formation of a ground state satisfying all of the pairwise interactions~\cite{balents2010}. The defining characteristics of frustration are massive ground-state degeneracy and concomitant strong fluctuations among these states. Thermal and quantum fluctuations suppress magnetic order and, under certain conditions, can lead to spin liquid regimes extending to low temperature. The honeycomb lattice is an interesting manifestation of a spin system where frustration arises from competing interactions rather than geometric constraints, and this frustration is further enhanced by strong quantum fluctuations due to the low coordination number z=3.

The system of interacting spins on a honeycomb lattice has attracted the attention of theorists for decades~\cite{rastelli1979,katsura1986}, with more recent calculations proposing that a spin liquid state can be stabilized on this lattice~\cite{meng2010, flint2013, gong2013, zhu2013}. Competition between first, second and third neighbor magnetic exchange interactions, J$_{1}$, J$_{2}$, and J$_{3}$, results in a rich magnetic phase diagram~\cite{rastelli1979,katsura1986}. For classical localized spins described by a Heisenberg Hamiltonian, N\'eel, stripy, zigzag and spiral magnetic orderings are possible depending on the relative strengths of these interactions. Further, three tricritical points, where three of these types of long range magnetic order become degenerate, are predicted and the strongest frustration would be expected near these points~\cite{rastelli1979}. 

The honeycomb lattice compounds MnTiO$_3$ and BaNi$_2$(PO$_4$)$_2$ were discovered early on~\cite{ishikawa1958, eymond1969}. Both were found to be N\'eel antiferromagnets with MnTiO$_3$ ordering at 64 K~\cite{shirane1959} and BaNi$_2$(PO$_4$)$_2$ ordering at 23.5 K~\cite{regnault1983}, in agreement with the determined exchange interactions that place them deep in the N\'eel phase of the theoretical honeycomb lattice phase diagram~\cite{todate1986, regnault1983}. More recently, there have been several experimental realizations of frustrated honeycomb lattice systems with antiferromagnetic interactions based on transition metals, e.g. Bi$_{3}$Mn$_{4}$O$_{12}$(NO$_{3}$) ~\cite{smirnova2009}, (Na/Li)$_{2}$IrO$_{3}$~ \cite{singh2010, liu2011,singh2012}, $\alpha$-RuCl$_3$~\cite{plumb2014} SrL$_{2}$O$_{4}$ (L = Gd, Dy, Ho, Er, Tm, and Yb)~\cite{karunadasa2005}, Cu$_{5}$SbO$_{6}$~\cite{climent2012}, and Cu$_{3}$M$_{2}$SbO$_{6}$ (M = Ni, Co)~\cite{roudebush2013}.  While inelastic neutron scattering measurements and complementary electronic structure calculations have placed bounds on the exchange interactions in  Bi$_{3}$Mn$_{4}$O$_{12}$(NO$_{3}$)~\cite{matsuda2010, kandpal2011, wadati2011} and Na$_{2}$IrO$_{3}$~\cite{choi2012, gretarsson2013}, determination of individual exchange interactions in these compounds has not been possible due to the lack of large single crystals and/or strong Ir absorption. Uncertainty remains over even the relative strength of the exchange interactions in these compounds. This has hindered comparison with theoretical phase diagrams, which propose spin liquid and highly frustrated phases depending on the value of the exchange interactions~\cite{meng2010, flint2013, gong2013, zhu2013, mazin2013}. 

We present inelastic neutron scattering results that characterize the exchange interactions in single crystals of the antiferromagnetic insulator {\CMS}, which consists of honeycomb layers of Mn in which every other atom is shifted perpendicular to the ab plane~\cite{cordier1976, simonson2012}. Neutron powder diffraction measurements reveal N\'eel-type antiferromagnetic order in {\CMS} below T$_{N}$ = 85 - 88 K, with an ordered moment between 2.8-3.4 $\mu_{B}$/Mn~\cite{ratcliff2009, bridges2009}. The  magnetic moment is substantially smaller than the 5 $\mu_{B}$/Mn expected from the high spin state produced by Hund's rules, and this reduced moment may reflect the interplay of quantum fluctuations and hybridization~\cite{simonson2012pnas, mcnally2014}. The moments are refined to lie in the honeycomb a-b plane, possibly with some degree of out-of-plane canting. Between T$_{N}$ and 210 K a weak ferromagnetic component was detected in magnetic susceptibility measurements~\cite{simonson2012}. From 340 K - 400 K, Curie-Weiss behavior was reported with a low paramagnetic moment of 1.4 $\mu_{B}$/Mn~\cite{simonson2012}. The low ordering temperature, as well as the unusual character of the intermediate temperature phase, suggest that frustration characteristic of the honeycomb lattice may be a crucial part of the magnetism of {\CMS}, unaddressed until now.

Our single-crystal inelastic neutron scattering measurements reveal spin wave excitations in {\CMS} at T = 5 K $\textless\textless$ T$_{N}$. We will show that these excitations are well described by a Heisenberg model of spins on a corrugated honeycomb lattice, allowing us to characterize the antiferromagnetic exchange interactions J$_{1}$ and J$_{2}$, as well as the exchange interactions between nearest neighbors in the c-direction J$_{c}$. Using the exchange interactions determined in this way, we situate {\CMS} on the theoretical magnetic phase diagram, and find it is proximate to a tricritical point, and is consequently magnetically frustrated.  

Our inelastic neutron scattering measurements were carried out on the SEQUIOA time-of-flight instrument at the Spallation Neutron Source at Oak Ridge National Laboratory~\cite{granroth2010}.   An incident energy of 50 meV was used yielding energy resolution of 1 meV, with the Fermi chopper 2 set to 420 Hz and the T0 chopper set to 90 Hz.   The measurements were performed on four co-aligned single crystals of {\CMS}, of total mass 3.2 g mounted on a sheet of aluminum in a displex helium closed cycle refrigerator.   These single crystals were grown from a Sn flux, as detailed elsewhere~\cite{simonson2012}.   

\begin{figure}[[htbp!]
\includegraphics[width=8.6 cm]
{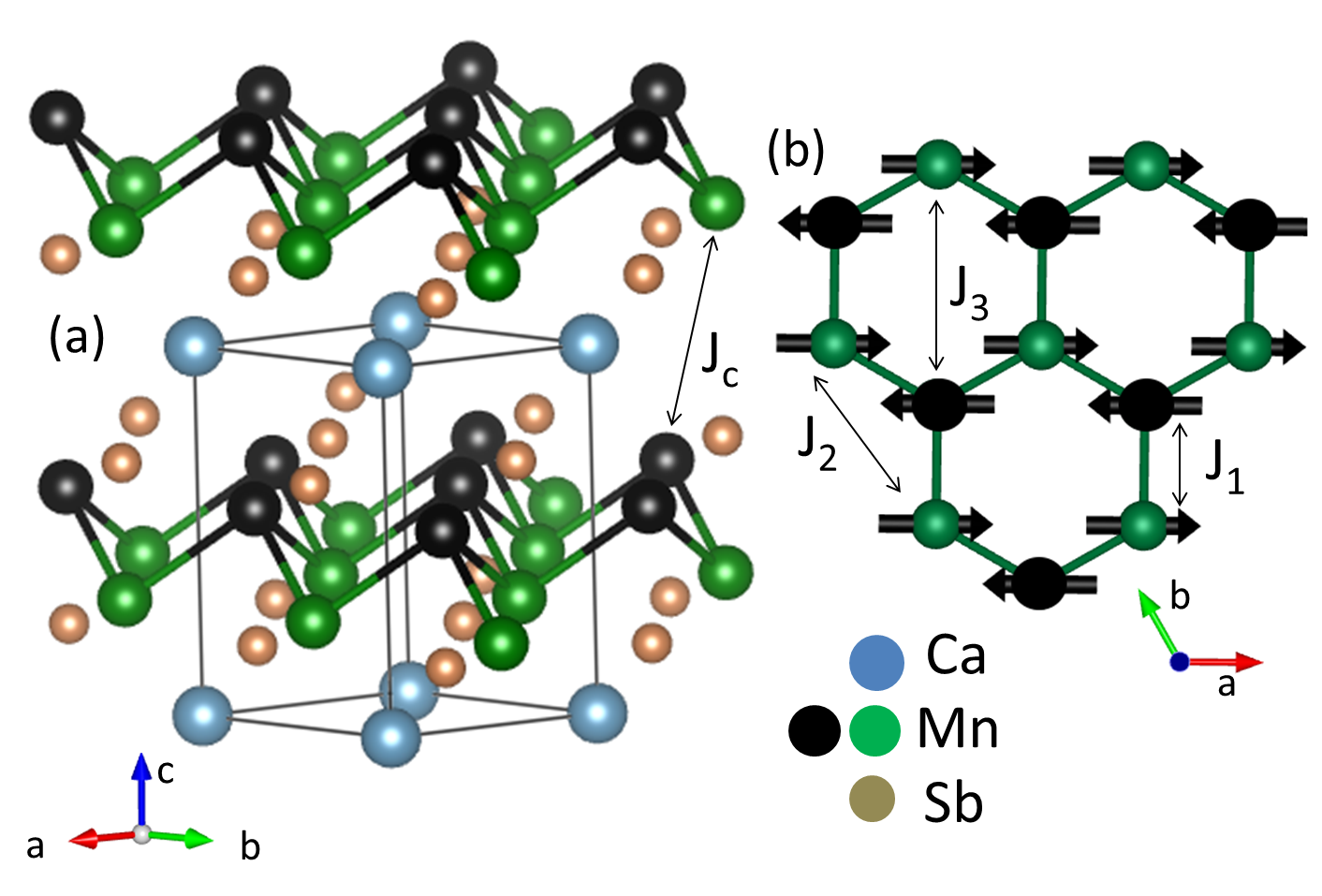}
\caption[]{(Color online) (a) The crystal structure of {\CMS}, which forms in the trigonal CaAl$_{2}$Si$_{2}$ structure type (S.G. P$\bar{3}$m1, no. 164). An outline of the unit cell is shown. The corrugation of the honeycomb layer of Mn is emphasized by the black and green Mn atoms displaced along c-direction. Exchange interaction along the c-direction J$_{c}$ is indicated. (b) A compressed view of the N\'eel antiferromagnetic corrugated honeycomb lattice formed by the Mn moments in the ab plane. Exchange interactions between first neighbors J$_{1}$, second neighbors J$_{2}$ and third neighbors J$_{3}$ are indicated.}
\label{Fig1}
\end{figure}

The crystal and magnetic structures of {\CMS} are presented in Figure~\ref{Fig1}. {\CMS} forms in the trigonal CaAl$_{2}$Si$_{2}$ structure type with lattice parameters \textit{a} = 4.52 {\AA} and \textit{c} = 7.46 {\AA}~\cite{bobev2006}. The corrugation between nearest-neighbor black and green Mn atoms in the ab plane is evident (Figure~\ref{Fig1}a). The first neighbor Mn-Mn spacing between respectively buckled Mn atoms is 3.2 {\AA} and the second neighbor spacing corresponds to the lattice constant \textit{a}. Most interestingly from a magnetic perspective, viewing the crystal structure from above (Figure~\ref{Fig1}b) reveals a honeycomb lattice of Mn. The magnetic moments associated with the Mn atoms form a N\'eel antiferromagnetic pattern in the ab plane below T$_N$ , and the nearest-neighbor exchange along the c direction J$_c$ is also antiferromagnetic (Figure~\ref{Fig1}a)~\cite{bridges2009,ratcliff2009}. We may consider the three first neighbor Mn spins as coupled by exchange interaction J$_{1}$ and six second neighbor spins coupled by exchange interaction J$_{2}$ (Figure~\ref{Fig1}b). These exchange interactions are mediated by one superexchange path connecting nearest neighbor Mn spins with $\angle$Mn-Sb-Mn angle of 70$^\circ$ and another connecting the second neighbors with $\angle$Mn-Sb-Mn = 108$^\circ$.

\begin{figure}[[htbp!]
\includegraphics[width=8.6 cm]
{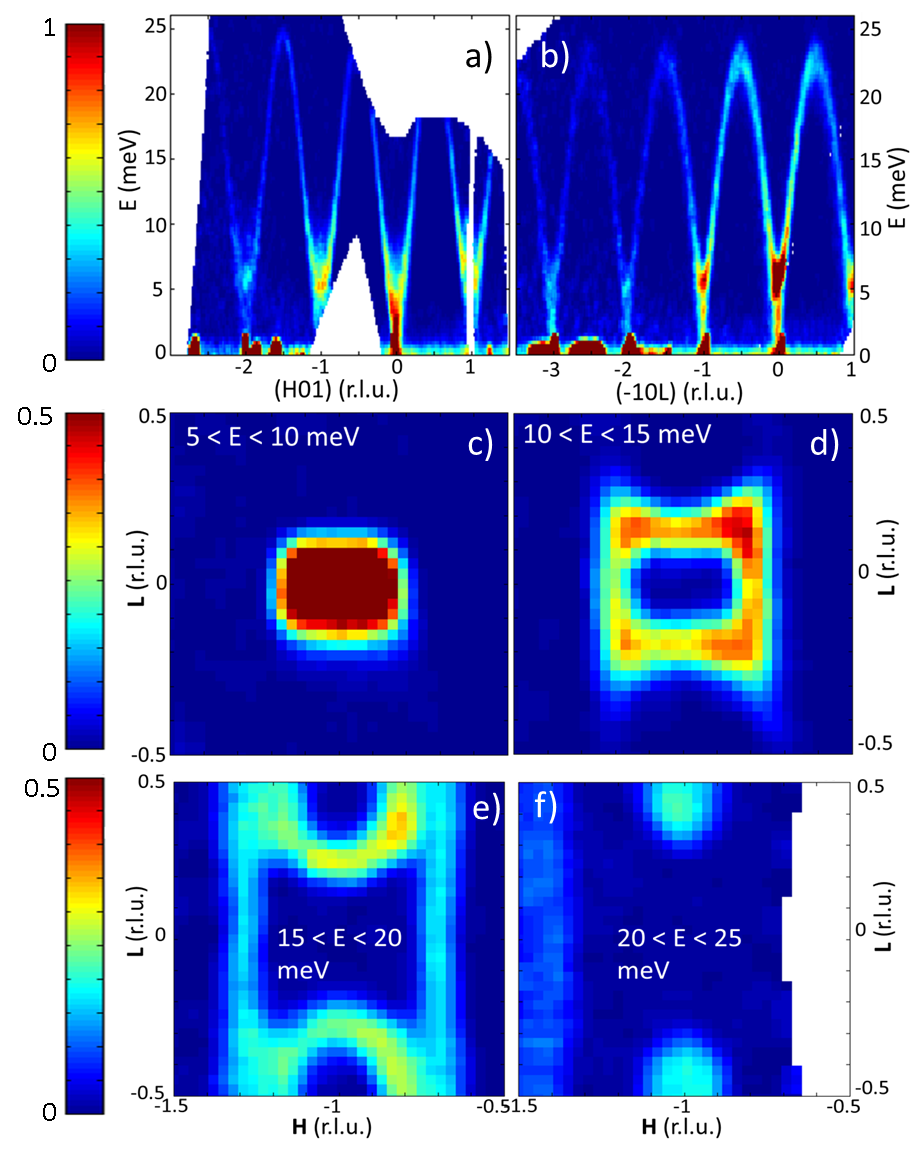}
\caption[]{(Color online) Contour plots of inelastic neutron scattering intensity at T = 5 K. Scale bars are shown on the left. Scattered neutron intensity as a function of energy E along the (a) \textbf{H} direction, (b) \textbf{L} direction. Scattered neutron intensity as a function of \textbf{H} and \textbf{L} for (c) 5 $\textless$ E $\textless$ 10 meV,  (d) 10$\textless$ E $\textless$ 15 meV,  (e) 15 $\textless$ E $\textless$ 20 meV,  (f) 20 $\textless$ E $\textless$ 25 meV}
\label{Fig2}
\end{figure}

Figure~\ref{Fig2} presents an overview of our inelastic neutron scattering measurements of CaMn$_{2}$Sb$_{2}$. The energy dependence of the scattered neutron intensity  S(\textbf{Q},E) along the \textbf{H} and \textbf{L} directions is presented in Figure~\ref{Fig2}a-b. Here, we define $\boldsymbol Q$ = \textbf{b}$_{1}$ h +  \textbf{b}$_{2}$ k +  \textbf{b}$_{3}$ l = (\textbf{H},\textbf{K},\textbf{L}) where \textbf{b}$_{1,2,3}$ are the reciprocal lattice vectors of the trigonal lattice~\cite{liu2012} Sharp, dispersive excitations emerge from all reciprocal lattice points with integer h and l values, as expected for spin waves in the N\'eel phase of a honeycomb lattice. Two spin wave branches are discernible corresponding to an acoustic mode and an optical mode emanating from the antiferromagnetic zone center and $\approx$ 4 meV respectively. The excitations are similar along the (H01) and (-10L) directions, with a maximum spin wave energy of $\approx$ 24 meV in both cases. S(\textbf{Q},E) is observed to decrease slightly as the wavevector increases, as expected from the magnetic form factor for Mn, acting in concert with the polarization dependent scattering from the ordered magnetic moments~\cite{brown, zaliznyak2005}.  Figures~\ref{Fig2}c-f present two-dimensional cuts along the \textbf{H} and \textbf{L} directions for increasing energy transfers. For data summed over energy  transfers 5 meV$\textless$E$\textless$10 meV (Figure~\ref{Fig2}c), we observe the most intense scattering in an oval shape centered at the Bragg position (h,k,l) = (-1,0,0). These data are consistent with the scattering expected from dispersive spin wave excitations. For larger energy transfers (Figure~\ref{Fig2}d-e) S(\textbf{Q},E) has only a two-fold rotational symmetry centered at the Bragg position, suggesting the spin waves disperse differently along the $\textbf{H}$ and $\textbf{L}$ directions. For 20 meV$\textless$E$\textless$25 meV (Figure~\ref{Fig2}f), the spin waves have dispersed to the edge of the Brillouin zone, consistent with a  magnon bandwidth of $\approx$ 25 meV. The energy and wavevector dependence of the scattered neutron intensity behaves just as expected for three-dimensional spin waves.

\begin{figure}[[htbp!]
\includegraphics[width=8.6cm]
{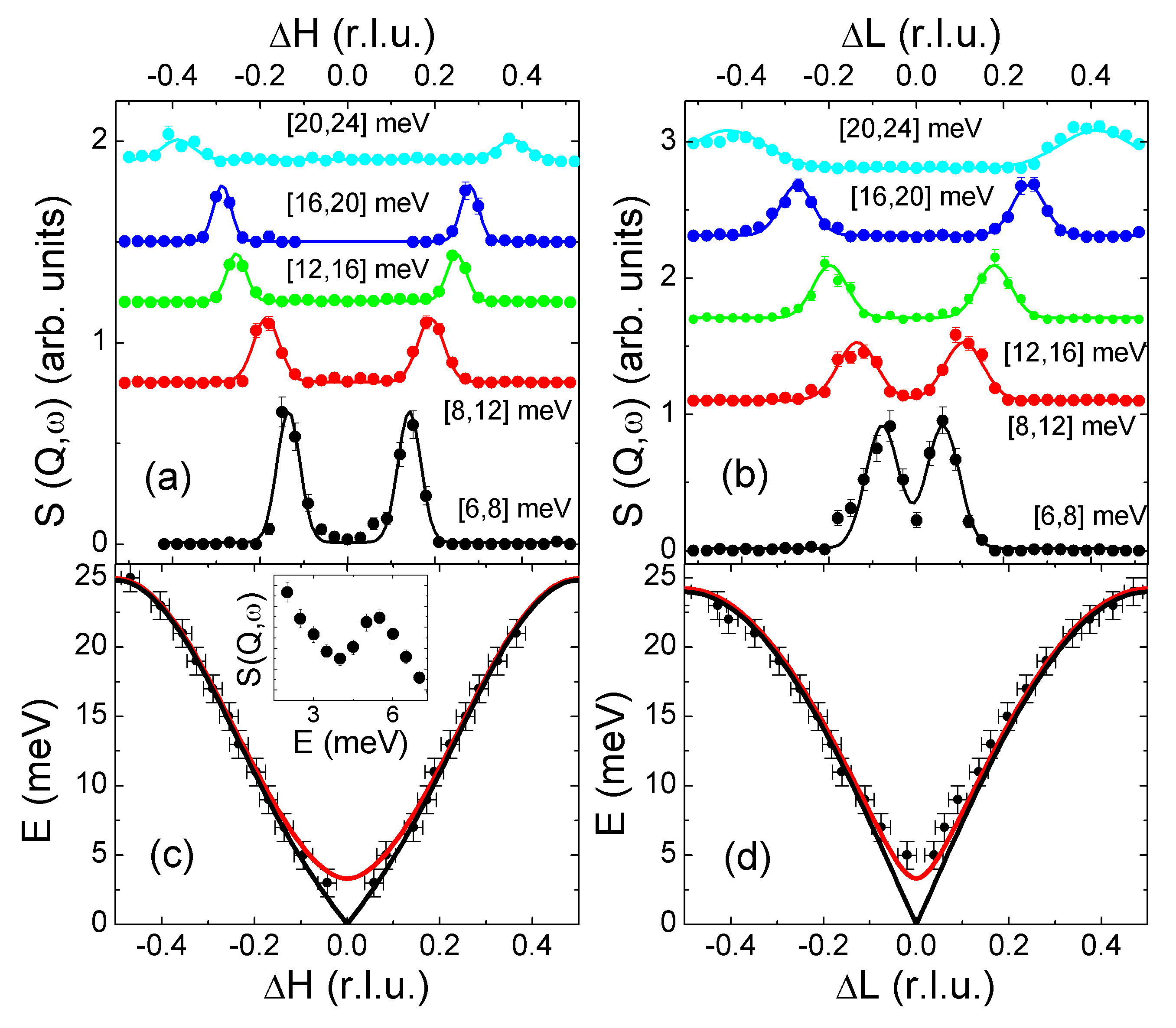}
\caption[]{(Color online) (a,b) Scattered neutron intensity S(\textbf{Q},E) along the \textbf{H} and \textbf{L} directions for ranges of energy transfers E, as indicated. Scans are displaced for clarity. The solid lines are fits to the measured data as described in the text. (c,d) Black points represent the spin wave momenta and energies along \textbf{H} and \textbf{L} extracted from fits. Solid lines are fits to the observed dispersion with that expected from a Heisenberg spin model, as described in the text. (inset) S(\textbf{Q},E) as a function of E at the zone center.}
\label{Fig3}
\end{figure}

In Figure~\ref{Fig3} we present fits to the observed scattering that allow us to extract the spin wave dispersion along the \textbf{H} and \textbf{L} directions, and to characterize the magnetic exchange interactions in {\CMS}. The scattered neutron intensity S(\textbf{Q},E) for different energy transfers along the \textbf{H} direction is shown in Figure~\ref{Fig3}a.  For summed energy transfers 6 meV $\leq$ E $\leq$ 8 meV, S(\textbf{Q},E) is well fitted by the sum of two Gaussian functions, shifted from the magnetic Bragg peak. At larger energy transfers the peak positions of the fits move further from the Bragg peak, just as expected for dispersing spin wave excitations. For E $\textgreater$ 24 meV we no longer observe scattering from the spin waves. Figure~\ref{Fig3}b presents S(\textbf{Q},E) along the \textbf{L} direction, where we again observe dispersive spin wave excitations. Fits along \textbf{H} and \textbf{L}, centered at the average spin momenta $\pm$$\Delta\textbf{Q}$(E), were performed every 2 meV . This fitting yields the spin wave momenta for different energy transfers, and the resulting spin wave dispersions, E($\Delta$\textbf{H}) and E($\Delta$\textbf{L}), are presented in Figures~\ref{Fig3}c,d. 

A Heisenberg model is used to fit the measured spin wave dispersion. Spin wave theory predicts that the one-magnon neutron scattering cross-section contains terms of the form~\cite{lovesey1984}

\begin{eqnarray}
 \frac{d^{2}\sigma}{d\Omega dE} \propto \sum_{\textbf{q},\boldsymbol\tau} (n_{\textbf{q}}+1) \delta(E({\textbf{q}}) - E) \delta(\boldsymbol\kappa-\textbf{q}-\boldsymbol\tau) 
\end{eqnarray}

where $\boldsymbol \tau$ are reciprocal lattice vectors for a single sublattice, $\boldsymbol \kappa$ is the wavevector transfer, $\boldsymbol q$ is the wavevector, n$_{\boldsymbol q}$ = (exp(E($\boldsymbol q$)/k$_B$T)-1)$^{-1}$, where k$_B$ is the Boltzmann constant. In linear spin wave theory, the Heisenberg Hamiltonian for an antiferromagnetic configuration of spins on a corrugated honeycomb lattice can be determined from the following dispersion relation:

\begin{equation}
E(\textbf{Q}) = 2S\sqrt{(J(0)-J'(0)+J'(\textbf{Q})+h_A)^2 - |J(\textbf{Q})|^2}
\end{equation}

Here, S is the total spin on an atom and h$_A$ is a reduced anistropy field. The exchange term J(\textbf{Q}) describes interactions betweens spins on opposite sublattices and J'(\textbf{Q}) describes interactions between spins on the same sublattice. The absolute value $|J(\textbf{Q})|$ must be taken because the honeycomb lattice is non-Bravais, like the diamond lattice~\cite{macdougall2011}. We include first neighbor exchange interactions J$_{1}$ for spins that are on opposite sublattices, second neighbor interactions J$_{2}$ for spins that are on the same sublattice, and exchange interactions between nearest neighbors in different honeycomb layers J$_{c}$ for spins that are on opposite sublattices. The interaction term J'(\textbf{Q}) = $\sum_{\textbf{n.n.n.}} J_{2} e^{i \textbf{Q}.\textbf{r}_{n.n.n.}}$, where the sum is over the 6 second neighbor atoms. The term J(\textbf{Q}) = $\sum_{\textbf{n.n.}} J_{1} e^{i \textbf{Q}.\textbf{r}_{n.n.}}$ + $\sum_{\textbf{c}} J_{c} e^{i \textbf{Q}.\textbf{r}_{c}}$.

The resulting expression for the spin wave dispersion for a J$_{1}$-J$_{2}$-J$_{c}$ Heisenberg model were fit simultaneously to the measured dispersions along the \textbf{H} and \textbf{L} directions. Fits were performed for a gapless acoustic mode (h$_A$ = 0) and a gapped optical mode and are shown respectively as black and red solid lines in Figure~\ref{Fig3}c-d. Excellent agreement is found between the Heisenberg model and the observed excitations. We find that SJ$_{1}$ = 7.9 $\pm$ 0.6 meV and SJ$_{2}$ = 1.3 $\pm$ 0.2 meV are both positive with J$_{2}$/J$_{1}$ = 0.165, signalling that the in-plane interactions are antiferromagnetic. The value of the ratio J$_{2}$/J$_{1}$ = 0.165 remains robust independent of the details of the microscopic model, that is whether or not the corrugation of the honeycomb planes or multiple anisotropy terms are included.  The exchange interaction betwen nearest neighbors in different honeycomb layers SJ$_{c}$ = 0.51$\pm$ 0.05 meV. The experimentally determined values of the exchange interactions are in good agreement with values obtained from density functional theory (DFT) calculations~\cite{mazin2013}. The values found in these calculations are SJ$_1$ = 13.5 meV, SJ$_2$ = 3.25 meV and SJ$_c$ = 0.45 meV.  DFT somewhat overestimates the exchange interactions as the Hubbard U was not included in the calculations. Introducing a third neighbor in plane exchange interaction J$_{3}$ or second neighbor out of plane exchange does not appreciably improve the accuracy of the model presented here, and indeed these terms were found to be small from DFT calculations~\cite{mazin2013}. Therefore we do not include these terms in our analysis and take J$_3$ = 0. The presence of a spin gap and gapped mode is confirmed in the inset to Figure~\ref{Fig3}c that presents scattered neutron intensity at the zone center as a function of energy. The anisotropy field h$_A$ opens a spin gap of 4 meV at the zone center and the effect of the competition between this anistropy and the exchange interactions on the phase diagram of the honeycomb lattice will be of interest for future theoretical calculations.

\begin{figure}[[htbp!]
\includegraphics[width=8.6cm]
{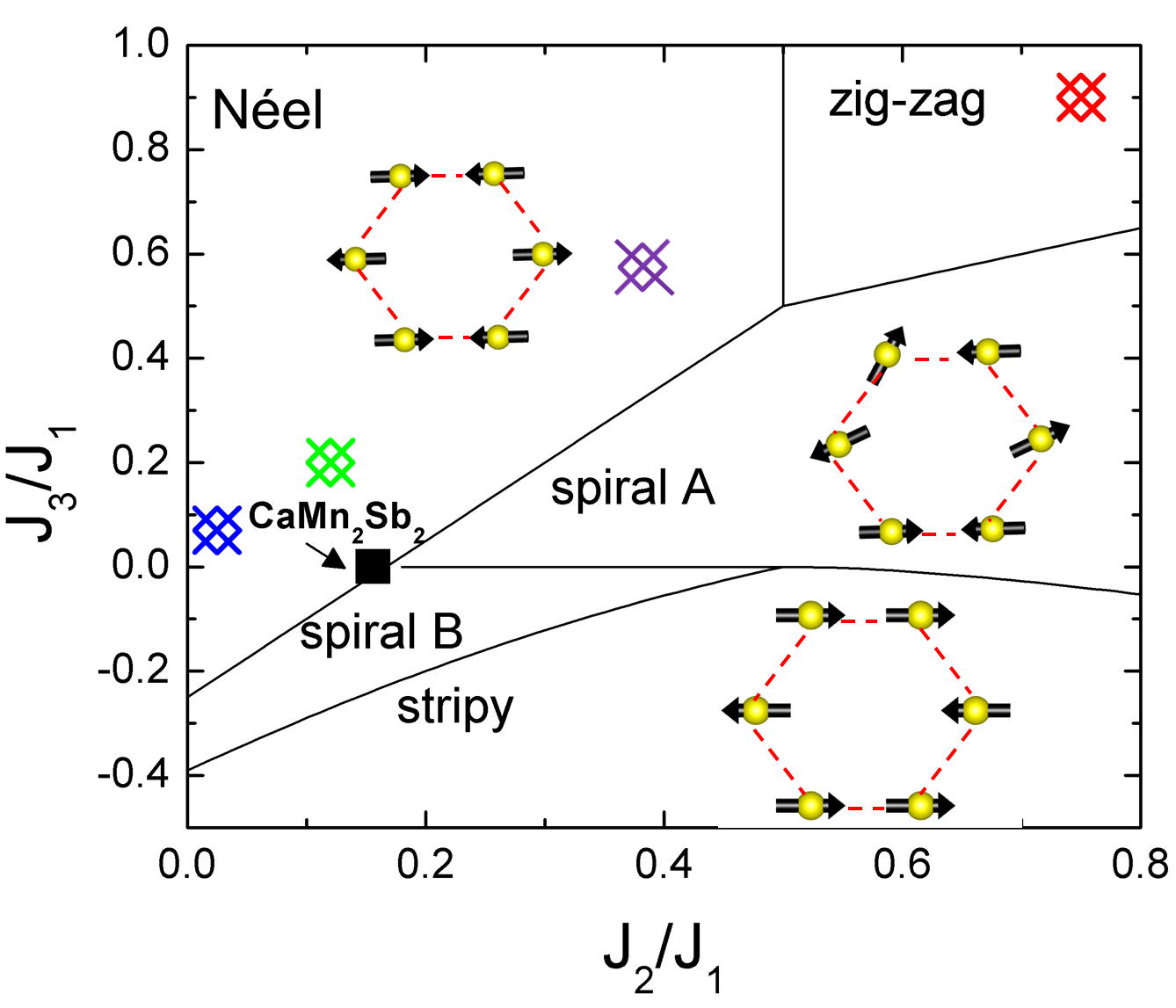}
\caption[]{(Color online) Phase diagram of the Heisenberg model for a honeycomb lattice with first, second and third neighbor exchange interactions J$_{1}$, J$_{2}$, and J$_{3}$ respectively~\cite{rastelli1979}. Solid lines are phase boundaries for the different antiferromagnetic configurations indicated. The blue, purple, green and red symbols represent MnTiO$_3$, BaNi$_2$(PO$_4$)$_2$, Bi$_3$Mn$_4$O$_{12}$ and Na$_2$IrO$_3$ respectively. Filled black square is {\CMS}}
\label{Fig4}
\end{figure}

In Figure~\ref{Fig4} we use the ratio of the experimentally determined exchange interactions to situate {\CMS} on the phase diagram of the classical J$_{1}$-J$_{2}$-J$_{3}$ Heisenberg model~\cite{rastelli1979} for a honeycomb lattice of spins, which is controlled by the ratios J$_{2}$/J$_{1}$ and J$_{3}$/J$_{1}$. Depending on the relative strengths of these interactions different types of antiferromagnetic ordering are expected, as indicated. Using the values of the exchange interactions determined from our inelastic neutron scattering measurements, we find that {\CMS} lies in the N\'eel ordered region of the phase diagram, in agreement with the magnetic structure determined from powder neutron diffraction measurements~\cite{ratcliff2009, bridges2009}. Further, {\CMS} is found to be very close to the tricritical point where N\'eel order and two spiral antiferromagnetic configurations are predicted to co-exist. This large degeneracy of possible ground states, as well as presumed strong fluctuations among these states, is likely responsible for the relatively low ordering temperature of {\CMS}, T$_{N}$ = 85 K~\cite{simonson2012, mcnally2014}, which is much reduced from the mean field ordering temperature T$_{MFT}$ = (S+1)(3SJ$_1$+6SJ$_2$+2SJ$_c$)/3k$_{B}$ = 310 K for S=3/2 or 370K for S= 2. The close proximity of {\CMS} to the tricritical point reported here confirms a recent prediction by Mazin~\cite{mazin2013}, who speculates that the weak ferromagnetic component found in the intermediate temperature range could result from this proximity. 

Until now, there has been a dearth of antiferromagnetic honeycomb lattice compounds whose exchange interactions have been determined experimentally, so as to facilitate comparison with the phase diagram in Figure \ref{Fig4}. The exchange interactions determined from a single crystal inelastic neutron scattering study of MnTiO$_3$ and BaNi$_2$(PO$_4$)$_2$ place them deep in the N\'eel phase~\cite{todate1986, regnault1983}, in agreement with the determined magnetic structure~\cite{shirane1959, regnault1983}. Bounds on the exchange interactions of the effective spin 1/2 honeycomb lattice compound Na$_2$IrO$_3$ place it solidly in the zigzag antiferromagnetic phase. While this is in agreement with the experimentally determined magnetic structure, a Kitaev exchange term is important to characterize the strong magnetic frustration in this compound, and the strong spin-orbit coupling may displace this compound from the indicated position~\cite{choi2012,gretarsson2013}. Inelastic neutron scattering measurements have also been reported on the honeycomb lattice compound Bi$_3$Mn$_4$O$_{12}$ and, using the resulting bounds on exchange interactions, this compound is also situated in the N\'eel antiferromagnetic phase of Figure \ref{Fig4}.  However, long range magnetic order in Bi$_3$Mn$_4$O$_{12}$ has not been observed down to 0.4 K, indicating interlayer exchange interactions are likely necessary to understand its magnetic properties~\cite{matsuda2010}.  Thus, our experiments are the first to show that {\CMS} is an antiferromagnetic honeycomb lattice compound situated in close proximity to a multicritical point on the phase diagram of the Heisenberg model for a honeycomb lattice. This proximity enhances the magnetic frustration and further reduces the ordering temperature in {\CMS} from the expected mean field ordering temperature. It would be interesting to study a structurally similar compound with stronger quantum fluctuations, e.g. by replacing the large Mn moments with lower spin moments, to determine if the long range magnetic order could be completely suppressed, leading to a spin liquid state.



We thank I.I. Mazin and S. Artyukhin for helpful discussions. We acknowledge the Office of the Assistant Secretary of Defense for Research and Engineering for providing the NSSEFF funds that supported this research. Work at BNL (I.A.Z.) was supported by Office of Basic Energy Sciences, US DOE, under Contract No. DE-AC02-98CH10886. Research conducted at the Spallation Neutron Source at ORNL was sponsored by the Scientific User Facilities Division, Office of Basic Energy Sciences, US Department of Energy.

\end{document}